# Effect of Self Diagnosis on Subsequent Problem Solving Performance

E. Yerushalmi[1], A. Mason[2], E.Cohen[1], C. Singh[2]

[1]Department of Science Teaching, Weizmann Institute of Science, Rehovot, Israel
[2]Department of Physics and Astronomy, University of Pittsburgh, Pittsburgh, PA 15213, USA

**Abstract.** "Self-diagnosis tasks" aim at fostering diagnostic behavior by explicitly requiring students to present diagnosis as part of the activity of reviewing their problem solutions. The recitation classes in an introductory physics class (~200 students) were split into a control group and three experimental groups in which different levels of guidance was provided for performing the self-diagnosis activities. We have been a) investigating how students in each group performed on subsequent near and far transfer questions given as part of the exams; and b) comparing student's initial scores on their quizzes with their performance on the exams, as well as comparing student's self-diagnosis scores with their performance on the exams. We discuss some hypotheses about the students' ability to self-diagnose with different levels of scaffolding support and emphasize the importance of teaching students how to diagnosis their own mistakes. Our findings suggest that struggling with minimal support during in-class self-diagnosis can trigger out-of-class self-diagnosis. Students therefore may be motivated to make sense of the problem they may have not been able to self diagnose, whether independently or in a collaborative effort.

## INTRODUCTION

In the previous two papers [1,2], we investigated what are students able to diagnose while reviewing their quiz problem solutions if deliberately prompted to self-diagnose. Table 1 [2] describes three alternative self diagnosis tasks differing in the instructions and resources students received from thorough to minimal scaffolding.

The results indicated that the students' ability to self-diagnose generally reflected the level of support. Students with a rubric and solution outline were able to self-diagnose the best, while students with a handout of the detailed solution performed moderately and students with only their text and notes performed the worst.

We were subsequently interested to investigate the impact of these results on the performance on a midterm problem that is isomorphic with the quiz problem that the students worked on during the recitation. The intent is to see if the self-diagnosis tasks will improve transfer from the quiz to the midterm exam.

Research focused on making analogies [3,4] shows that many students don't know how to use a worked out example to solve a transfer problem (similar in required general procedure (principles/intermediate variables), different in detailed procedures (surface features)). Students' representation that is organized around surface features [5,6] prevents students from retrieving and implementing procedures from the worked out example. Medium and high achieving students benefited most from instruction explicitly presenting them with the procedures and a worked out examples rather than merely worked out examples [4]. Similarly we hypothesized that diagnosing one own solutions using solution outline and a rubric that focus his/her attention on the procedure will enhance transfer to isomorphic problems.

The effectiveness of the tasks will be analyzed via success in transfer (isomorphic) problems both in terms of retaining the corrected ideas (i.e. invoking and applying appropriately the physics principles and concepts required), and in retaining characteristics of the presentation of reasoning, justifying the solution in a manner reflecting a strategic problem solving approach.

## DATA COLLECTION

It was our intention to achieve ecological validity, namely, to simulate conditions that are feasible in actual implementation in a classroom given the time constraints of teachers and students. Consequently we performed the experiment in actual classrooms and accepted the modifications introduced by the instructors who participated in the experiment. The study involved an introductory algebra based course for pre-meds (N~200), one instructor and two teaching assistants. TA classrooms were distributed into control groups and three self-diagnosis treatments groups who each carried out a different self-diagnosis task (see table 1). In all treatment groups students first solved a quiz problem during a recitation and were asked in the next recitation to circle mistakes in their photocopied solutions and explain what they did wrong.

**TABLE 1:** Distribution of groups among SD tasks

|  | self-diagnosis tasks | | |
|---|---|---|---|
| *Group A* | groups (B) | group (C), | 1 group (D), |
| control | Instructor outline, diagnosis rubric | Worked out example | Minimal guidance: notes + text books |
| ~100 students *3 sections* | 31 students | 25 students | 24 students |

An outcome of the decision to perform the experiment in actual classrooms is that we must consider the effect of the differences in TAs and the effect of interactions within the groups. It is possible that the teaching assistants' different styles as well as the interaction within each group have introduced differences in performances of different groups. Both intergroup (between-subjects) and intragroup (within-subjects) effects will be examined. We will further compare each TA's groups separately so that the difference between TA styles is not relevant. One TA presided over groups C and D and the other TA was in charge of group B and the control group.

## IMPLICATION OF IMPLEMENTATION IN ACTUAL CLASSRROMS

As the study is performed in the context of actual calssroom, there are two main kinds of potential learning processes of interest in which the learner elicits knowledge required to transfer to a similar problem. The first kind takes place while performing the self diagnosis task and is primarily individual. The second kind of learning, which can be cooperative as well as individual, is a subsequent manifestation of the diagnostic activity that might provoke a sort of uneasiness; the student becomes aware that he does not feel comfortable with his current knowledge level. This can promote group as well as individual dynamics that promote additional subsequent learning.

There was less than a week on average between the self diagnosis task in which students reflected on past attempts at a quiz problem and the midterm. During this period of time, the solution to the problem was posted on Courseweb. Students could look at the solution to the quiz problem on their own before the midterm and continue analyzing the problem as well as discuss with each other how the problem was done. This required further considerations when dealing with the impact of self-diagnosis on the midterm.

## THE MIDTERM PROBLEM

**FIGURE 1.** Midterm problem.

> A family decides to create a tire swing in their back yard for their son Ryan. They tie a nylon rope to a branch that is located 16 m above the earth, and adjust it so that the tire swings 1 meter above the ground. To make the ride more exciting, they construct a launch point that is 13 m above the ground, so that they don't have to push Ryan all the time. You are their neighbor, and you are concerned that the ride might not be safe, so you calculate the maximum tension in the rope to see if it will hold. (a) Where is the tension greatest? (b) Calculate the maximum tension in the rope, assuming that Ryan (mass 30 kg) starts from rest from his launch pad. Is it greater than the rated value of 750 N? (c) Name two factors that may have been ignored in the above analysis, and describe whether they make the ride less safe or more safe.

The problem used in the midterm within a week after the self-diagnosis exercise is described in figure 1. This problem is similar to the quiz problem in that it employs the same physical principles, i.e. Newton's second law applied in a non-equilibrium situation involving centripetal acceleration and conservation of mechanical energy. Thus the solution will be isomorphic to the quiz solution.

This is illustrated in Table 2. We expect this problem to be as isomorphic with the quiz problem because the principles needed to solve either problem are the same. Furthermore, both questions also require recognition of similar target variables (in the form of either a normal force or tension force) and intermediate variables (centripetal acceleration and velocity at the maximum point on a circular trajectory). They differed in terms of context as well as what the direction of centripetal acceleration was with respect to the normal force.

**TABLE 2:** Comparison of quiz and midterm problems.

|  | Princ-iples | Varia-bles | FBD | Context | details |
|---|---|---|---|---|---|
| Quiz | EC $2^{nd}$ law | v $a_c$ N/T | ↑ N/T ↓ $F_g$ | Roller coaster | $a_c$ ↓↑ N |
| Midterm | | | | Tire swing | $a_c$ ↑↑ T |

As before, we are interested in the effect of the quiz diagnosis on the students' ability to solve the transfer problem as well as on the presentation of the reasoning when solving the problem. .

# RESULTS

As shown in the previous two papers [1,2] we differentiated the researcher's judgment of the students' self-diagnosis and solution into physics and presentation grade.

## Physics

Table 3 shows the mean physics scores for all physics groups on the midterm problem. To be able to consider the effect of the TAs on the inter-group comparison we present analysis of each TA's groups separately. Table 4 shows ANCOVA p-value comparisons between group B and the control group and between group C and group D, respectively. They show that indeed group B, that was provided with a rubric and solution outline, did significantly better than the control group A. One might conclude that that the rubric and solution outline provided students with a clear picture of what they did wrong.

**TABLE 3.** Means and standard deviations of each group's midterm physics scores.

|           | First TA |       | Second TA |       |
|-----------|----------|-------|-----------|-------|
| Group     | A        | B     | C         | D     |
| Mean      | 0.424    | 0.526 | 0.329     | 0.473 |
| Std. Dev. | 0.042    | 0.053 | 0.048     | 0.063 |

**TABLE 4.** P-value comparison between midterm physics scores of each TA's groups.

| First TA | Group B |
|----------|---------|
| Group A  | 0.112   |

| Second TA | Group D |
|-----------|---------|
| Group C   | 0.071   |

**TABLE 5.** Correlation of physics scores: self-diagnosis vs. midterm.

| Group | Correlation | p value |
|-------|-------------|---------|
| B     | 0.35        | 0.16    |
| C     | 0.13        | 0.55    |
| D     | 0.07        | 0.80    |

However, despite actually seeing the complete solution, group C fared worse on the midterm than group D, who did not receive the solution during the recitation but had to try and figure it out on their own. There are at least two possible interpretations for this result. First, it is apparent that group C's self-diagnosis was not meaningful in the sense that an elaborated solution allows self-diagnosis to occur on a more superficial level. Students can simply compare and contrast their answer with the detailed correct solution, and not necessarily think deeply about what they are doing wrong. Second, group D's relatively better performance can be understood if there was a self-diagnosis stage that occurred after the formal self-diagnosis exercise. That is, after the students struggled twice to solve the problem without the explanation provided to the other two groups, group D may have tried to make sense of the solution provided to all students after the exercise concluded. This may have led to a more thorough understanding of the problem providing them with deeper insight into the correct solution. In essence, group D may have actually received a more valuable self-diagnosis treatment than group C. While both groups had access to the solution at some point, group D students also had prior confrontation with their inability to solve and diagnose the problem, but group C's experience was largely self-contained to the solution.

The intra-group comparison is shown in table 5 in the form of a correlation between the self-diagnosis physics scores of the treatment groups and their midterm scores. A minor positive correlation for group B's self-diagnosis score is shown, indicating a tendency for the self-diagnosis to have helped students somewhat on the midterm. All other intra-group comparisons yield no correlations. This might also be explained in that between the self diagnosis task and the midterm, a separate learning process takes place.

## Presentation

Table 6 gives the mean presentation scores for all groups on the midterm problem, and an inter-group analysis of p-values between the treatment groups are shown in table 7. There is no significant difference or correlation between any of the groups. In addition, there was no effect of the treatment on presentation performance, even though group B did better on the self-diagnosis exercise. An intra-group analysis, presented in table 8, shows no correlation of the midterm scores with self-diagnosis scores

Since group B fared as poorly as the other groups with regard to presentation score on the midterm despite a better performance on presentation self-diagnosis, we must consider the "one-time" nature of the intervention. The students only performed the self-diagnosis exercise once, which would probably not be enough to effectively develop presentation skills, even though it is possible to understand the physical principles necessary to solve the problem. It would be of interest to examine whether applying the intervention consistently throughout the semester would help develop presentation skills.

**TABLE 6.** Means and standard deviations of each group's midterm presentation scores.

|  | First TA | | Second TA | |
|---|---|---|---|---|
| Group | A | B | C | D |
| Mean | 0.426 | 0.437 | 0.410 | 0.462 |
| Std. Dev. | 0.020 | 0.025 | 0.022 | 0.030 |

**TABLE 7.** P-value comparison between midterm presentation scores of each TA's groups.

| First TA | Group B | | Second TA | Group D |
|---|---|---|---|---|
| Group A | 0.580 | | Group C | 0.154 |

**TABLE 8.** Correlation of presentation scores: self-diagnosis vs. midterm.

| Group | Correlation | p value |
|---|---|---|
| B | -0.208 | 0.409 |
| C | -0.162 | 0.482 |
| D | 0.081 | 0.791 |

## DISCUSSION

We have described relative differences between the groups on both physics and presentation scores on the midterm. With regard to objective scoring, however, scores are poor in both categories for all groups ($0.329 <$ physics average $< 0.526$, $0.410 <$ presentation average $< 0.437$). The reason that none of the groups did well on the midterm can stem either from the task providing less scaffolding than necessary, or from the fact that the midterm problem was a far rather than near transfer problem as we originally expected.

The scaffolding involved a-priori modeling; the TA demonstrated how he would perform the diagnostic task on a mistaken solution he provided the students. Yet the coaching activity lacked feedback on how well students performed the diagnosis. Indeed the scaffolding was limited, meeting the time constraints of TAs in a large college classroom.

If transfer occurs, the student will recognize that the solution will require the same physics principles as the solution of the original quiz problem. The student will realize that the tension force in the rope is analogous to the normal force of the track on the roller coaster in the quiz problem, except here, the centripetal acceleration is in the same direction as the tension force, as opposed to the quiz problem in which the centripetal acceleration opposed the normal force. However, the student must first recognize that the maximum tension on the rope is at the lowest point of the tire's path. If the student does not realize this last fact, the resulting representation will no longer be analogous to the situation described by the quiz problem and the student cannot be expected to transfer the obtained knowledge. This problem is therefore a "far" transfer problem.

We have performed an independent categorization study (unpublished) in which we asked introductory-level students to group together 25 problems based on similarity of the material. Two of the problems in this study were exactly the midterm problem and the quiz problem that was the subject of self diagnosis in the study featured in this paper. Our research shows that students did not understand the similarity of the two problems and often placed them in different, mutually exclusive categories. For example, the midterm problem, which talks about a child on a tire swing, was often placed exclusively in a "pendulum" category or "tension" category that the quiz problem would not be placed in as it does not have a pendulum or deal with a tension force. If students at the introductory level did not even associate these two problems in the same group, it is probable that the midterm problem here was too far of a transfer for students to overcome.

We propose to repeat the analysis on another set of near transfer paired problems with the belief that students will perform better on the midterm problem overall after the self-diagnosis exercise with near transfer than with far transfer.

## ACKNOWLEDGMENTS

The research for this study was supported by ISF 1283/05 and NSF DUE-0442087.

## REFERENCES

1 A. Mason, E. Cohen, E. Yerushalmi, and C. Singh *Proceedings of the* Phys. Ed. Res. Conference, Edmonton, AIP Conf. Proc. Melville New York **1064**, 147, (2008).
2. E.Cohen, A. Mason, C. Singh and E. Yerushalmi *Proceedings of the Phys. Ed. Res. Conference, Edmonton,* AIP Conf. Proc. Melville New York **1064**, 99, (2008).
3. M. Gick, & K. Holyock, *Cognitive Psychology*, 15, 1-38 (1983).
4. B. Eylon and J. Helfman, In A.M. Mayer and P. Tamir (Eds.), *Science Teaching in Israel: Origins, Development and Achievements* . Jerusalem: Israel Science Teaching Center, 259-271. (in Hebrew) (1984).
5. M.T.H. Chi, P.J. Feltovich and R. Glaser, Cognitive Science, 5, 121-152 (1981).
6. B. Eylon and F. Reif, Effects of knowledge organization on task performance. *Cognition and Instruction*, 1(1), 5-44 (1984).
7. E. Yerushalmi, C. Singh, and B. Eylon, *Proceedings of the Phys. Ed. Res. Conference, Greensboro, NC,* AIP Conference proceedings, Melville NY **951**, 27-30, (2007).
8. C. Singh, E. Yerushalmi, and B. Eylon, *Proceedings of the Phys. Ed. Res. Conference, Greensboro, NC,* AIP Conference proceedings, Melville NY **951**, 31-34, (2007).